\documentclass[journal=jctcce,layout=manuscript,manuscript=article]{achemso}

\usepackage[version=3]{mhchem} 
\usepackage{booktabs}
\usepackage{graphicx}
\usepackage[table,xcdraw]{xcolor}
\usepackage{chemformula} 
\usepackage[T1]{fontenc} 
\usepackage{lscape}
\newcommand{\markNM}[1]{{\color{black} #1}} 

\author{Nisha Mehta}
\author{Golokesh Santra}
\author{Jan M. L. Martin}

\email{gershom@weizmann.ac.il}
\phone{+972-8-9342533}
\fax{+972-8-9343029}
\affiliation[Weizmann Institute of Science ]
{Department of Molecular Chemistry and Materials Science, Weizmann Institute of Science, Re\d{h}ovot, Israel}

\title[An \textsf{achemso} demo]
  {Is explicitly correlated double hybrid DFT advantageous for vibrational frequencies?}

\usepackage{xcolor}
\usepackage{etoolbox}
\usepackage{verbatim}
	
\usepackage{soul}
\newcommand{\wavenum}{{cm$^{-1}$} } 
\begin{document}



\begin{abstract}
We have investigated the effect of F12 geminals on the basis set convergence of harmonic frequencies calculated using two representative double-hybrid density functionals, namely B2GP-PLYP and revDSD-PBEP86-D4. Like previously found for energetics [N. Mehta and J. M. L. Martin, \textit{J. Chem. Theor. Comput.} \textbf{18}, 5978--5991 (2022)] one sees an acceleration by two zeta steps, such that even the cc-pVDZ-F12 basis set is quite close to the complete basis set (CBS) limit. However, the basis set convergence problem is not as acute as for energetics, and compared to experimental harmonic frequencies, conventional orbital calculations with augmented triple zeta quality basis set are acceptably close to the CBS limit, and can be carried out using analytical second derivatives. An efficient implementation of double hybrid-F12 analytical derivatives would make the F12 approach attractive in the sense that even an $spd$ orbital basis set would be adequate. For the accurate revDSD-PBEP86-D4 functional, the role of differing local correlation terms (Perdew-Zunger 1981 vs. VWN5) in different electronic structure programs has been investigated: while optimal double hybrid parameters and performance statistics for energetics as well as frequencies differ slightly between the two implementations, these differences are insignificant for practical purposes.
\end{abstract}



\section{Introduction}

Among the various ground-state molecular properties of a molecule that can be calculated through wavefunction ab initio methods (WFT) or density functional theory (DFT), harmonic frequencies are second in importance only to reaction energies and geometries. This status does not just derive from their being the starting point for the simulation of infrared and Raman spectra (with anharmonicity accounted for either through higher derivatives, or very approximately through scaling, see Ref.\cite{jmlm260} and references therein). They are also required for zero-point and thermal corrections to reaction energies (again, with some approximate account for vibrational anharmonicity), as well as to characterize stationary points on the potential energy surface as minima, transition states, or higher-order saddle points.

Harmonic frequencies of molecules cannot be observed directly. However, for diatomics and small polyatomics, if enough overtone and combination bands are observed, a power series in the vibrational quantum numbers can be established, from which `experimental' harmonic frequencies can be extracted (see, e.g., Csaszar\cite{Csaszar2012} for a review). More recently,  as illustrated by the work of, e.g.,  Jensen, Cs\'asz\'ar, Tennyson, Bowman, Puzzarini, Barone, and coworkers\cite{MORBID,TROVE,MARVEL,Csaszar2007,Bowman2009,Melli2021} (see also Romanowski, Gerber, and Ratner\cite{Ratner1988} for a very early example), spectral inversion techniques enable the `reverse engineering' of anharmonic force fields (and hence, of their harmonic component) from collections of observed spectral lines of one or more isotopomers of the molecule. A related `semi-experimental' approach, the term being first introduced for geometries\cite{JeanDemaison}, involves simplifying the potential inversion by setting most or all anharmonic parameters to calculated values, while fitting the harmonic (and perhaps a few key anharmonic) parameters to the observed band origins. (See Refs.\cite{Miani2000,Miani2001,jmlm135,jmlm172} and references therein for some early examples.)

Some of the challenges of accurate computational \markNM{vibrational} spectroscopy are  illustrated by a recent benchmark study on formaldehyde by Schaefer and coworkers\cite{Morgan2018}. As shown, e.g., by Karton and Martin\cite{jmlm230} (see also Ref.\cite{jmlm117}), the good performance of valence CCSD(T)\markNM{\cite{qcisdt_ccsdt,Watts1993}} for harmonic frequencies is actually the result of an error compensation: for diatomic hydrides, this takes place between neglect of inner-shell correlation (which pushes frequencies up) and neglect of higher-order triple excitation effects (which pull frequencies down). For heavy-atom diatomics, full CCSDT\markNM{\cite{Noga1986}} with valence correlation will actually lead to {\em over}estimated frequencies, and only connected quadruple excitations will slightly reduce curvature and bring RMS error down to the 2 \wavenum range, with a further reduction to 1 \wavenum from inclusion of connected \textit{quintuples}.\cite{jmlm230}

Ref.\cite{jmlm258} contains a compilation of harmonic frequencies of closed-shell diatomics and small polyatomics denoted HFREQ2014. In the later Ref.\cite{jmlm260}, the open-shell diatomics S$_2$ and SO were added, and three molecules were omitted (F$_2$ on account of severe static correlation, HNO on account of very high anharmonicity, and CF$_2$). In the present work, we shall use the Ref.\cite{jmlm260} dataset minus the two open-shell diatomics, which we shall denote HFREQ25.

As shown in Refs.\cite{jmlm258,jmlm260} for the said dataset, CCSD(T) near the basis set limit can achieve RMSD=5 cm\markNM{$^{-1}$}. For B3LYP\markNM{\cite{b3lyp_1993_Becke,b3lyp_2_1994}}/def2-QZVPD\markNM{\cite{def2basen,def2nZVPPD}}(very close to the basis set limit), one obtains\cite{jmlm260} RMSD=25.4 \wavenum, slightly better than the MP2 value of 30 \wavenum. 

Double hybrids, however, can improve on that. B2GP-PLYP\markNM{\cite{B2GPPLYP}}/aug-cc-pV(T+d)Z\markNM{\cite{dunning_1989, dunning_1992, dunning_2001}} reaches RMSD=10.7 \wavenum upon scaling frequencies by 0.992, while the spin-component-scaled double hybrid DSD-PBEP86-D2\markNM{\cite{jmlm238,jmlm251}}/aug-cc-pV(T+d)Z reaches the same essentially unscaled (optimal scale factor 0.9993). Analytical second derivatives\cite{refDHhessiansGaussian}, or seminumerical Hessians from finite differences of analytical first derivatives\cite{refDHgradORCA}, are available for double hybrids, making them an attractive alternative to both lower-rung DFT options (in terms of accuracy) and to coupled cluster theory (in terms of drastically reduced cost). For perspectives, see, e.g., Refs.\cite{Barone2021,Bloino2021}.

Alas, the GLPT2 (G\"orling-Levy 2nd-order perturbation theory\cite{Gorling1994}) term in double hybrids inherits the slow $L^{-3}$ basis set convergence\cite{Schwartz1962,kutzelnigg_Morgan1992} of MP2, $L$ being the highest angular momentum in the basis set. Recently, two of us have shown\cite{jmlm314,jmlm318} for the very large and chemically diverse GMTKN55 benchmark suite\cite{gmtkn55} how, for energetic properties, this can be overcome through the use of F12 geminals\cite{Ten-no2004} in the GLPT2 term.

This prompts the question whether DH-F12 (double hybrids with F12) offers a way out of the conundrum for harmonic frequencies as well, and this is the research question we are addressing in the present contribution. The answer should in principle be affirmative, as basis set convergence for frequencies tends to be faster\cite{jmlm053} than for energetics. At present, no analytical derivative implementation exists for DH-F12; for small to medium molecules, the cost of needing to evaluate the force constant matrix by numerical differentiation is compensated to some degree by the `embarrassingly parallel' character (and hence perfect parallelization) of the procedure. 


\section{Computational details}

All double hybrid-F12 calculations, and the orbital-only double hybrid calculations, were carried out using MOLPRO 2022\cite{molpro_2020} running on the Faculty of Chemistry HPC facility `ChemFarm' at the Weizmann Institute. Gradients and Hessians for these calculations were obtained by numerical differentiation of energies, using a stepsize of 0.005 bohr and a Krack-K\"oster\cite{Krack1998} integration grid of order 10 and using a cutoff of 10$^{-24}$. Global thresholds were set to 10$^{-12} E_h$ for energy convergence, 10$^{-11}$ for orbital convergence, 10$^{-18}$ and 10$^{-20}$ for individual two-electron integrals and shell prefactors, respectively. Geometries were optimized to a maximum displacement tolerance of 10$^{-5}$ bohr.

Some calculations were repeated using Gaussian 16 rev. C.01\cite{gaussian} with the \texttt{opt=Tight Grid=SuperFineGrid} options and analytical second derivatives. In doing so with identical basis sets and functionals, we found a discrepancy of just 0.3 \wavenum RMS between double-numerical and analytical values.

Finally, harmonic frequencies for the Berkeley functionals\markNM{\cite{wB97XV,wB97MV,wB97M2}} were obtained using Q-CHEM 6.0.1\cite{q-chem} with an unpruned grid of 150 radial Euler-Maclaurin\cite{Murray1993} and 590 angular Lebedev\cite{Lebedev1992,Beentjes} points.

We considered different families of basis sets. 
The first category are the correlation consistent \markNM{basis set of Dunning and co-workers},\cite{dunning_1989, dunning_1992, dunning_2001} which were developed with orbital-based correlated wavefunction calculations in mind (optimized for CISD valence correlation energies of atoms).
The notation haVnZ+d (n=D, T, Q, 5), in this paper, is shorthand for the combination of cc-pVnZ on hydrogen, aug-cc-pVnZ on first-row elements, and aug-cc-pV(n+d)Z on second-row elements.\par 
The second class of basis sets, purpose-developed for explicitly correlated calculations, are the cc-pVnZ-F12 (abbreviated VnZ-F12 in this manuscript) of Peterson and co-workers,\cite{peterson2008} or their anionic-friendly variants aug-cc-pVnZ-F12 (AVnZ-F12).\cite{martin_opus_276}
In the F12 geminal, the Slater exponent ($\beta$) values of 0.9, 1.0 and 1.0 were used for the (A-)VDZ-F12, (A-)VTZ-F12, and (A-)VQZ-F12, respectively. \par 
Finally, we also considered the Weigend-Ahlrichs/Karlsruhe def2 family,\cite{def2basen} namely def2-TZVPP and def2-QZVPP, and their diffuse function-augmented variants def2-TZVPPD and def2-QZVPPD.\cite{def2nZVPPD} \par
 
\subsection{A remark on the revDSD-PBEP86 local correlation component}

Concerning revDSD-PBEP86-D4, one ought to be aware that the implementations of the P86c correlation functional\cite{Perdew1986} in various codes are not necessarily equivalent. Or, to clarify, the enhancement factors/`gradient corrections' may be equivalent, but different codes apply them to different LDA correlation functionals. In particular, Gaussian 16 uses Perdew-Zunger 1981\cite{PerdewZunger1981} by default, which is not implemented in MOLPRO nor in ORCA (which, somewhat anachronistically, uses PW92LDA by default). MOLPRO uses VWN5,\cite{Vosko1980} which can be invoked \markNM{through} nonstandard inputs in both Gaussian (using \texttt{iop(3/74=1018)}) and ORCA (using \texttt{LDAOPt=C\_VWN5}).

When recalculating the Gaussian analytical frequencies using VWN5-P86c instead, we found discrepancies of several \wavenum between the two versions. It occurred to us that the published parameters\cite{revDSDfuncMartin} were obtained using the Perdew-Zunger LDA correlation\cite{PerdewZunger1981} which is the default in Q-Chem (used for most of that paper) as well as Gaussian. This prompts the question \markNM{of} whether substituting VWN5 or PW92LDA in the parametrization process might actually lead to slightly different parameters. In order to verify this, we evaluated all components of the GMTKN55 dataset in the exact same manner as detailed in Ref.\cite{revDSDfuncMartin}, except that, as \markNM{has been} our practice since Ref.\cite{jmlm299}, we also applied the def2-QZVPPD basis set to the BH76  \markNM{barrier heights subset,\cite{htbh38,nhtbh38,GMTKN24}} which contains some anionic S$_N$2 reactions. The original and revised parameters can be found in Table~\ref{tab:wtmad2}.

\begin{table}[h]
\scriptsize{
    \begin{tabular}{l|c|cccccc|cc}
     \hline
       ~~~ &LDA corr.&\multicolumn{5}{c}{parameters}&& \multicolumn{2}{c}{WTMAD2 of GMTKN55}\\
       Functionals &  part of &\multicolumn{6}{c}{~~}& opt param. & other param.\\
       ~~~  & of P86c & $c_{C,DFT}$ & $c_{2ab}$ & $c_{2ss}$ & $s_{6}$  & $a_{1}$ & $a_{2}$ & \multicolumn{2}{c}{(kcal/mol)}\\
     \hline
        revDSD-PBEP86-D3BJ & PZ81 & 0.4312 & 0.5761 & 0.0816 & 0.4339 &  0.0 & 5.5 & 2.37 & 2.37\\
        revDOD-PBEP86-D3BJ && 0.4443 & 0.6047 & 0.0 & 0.4790 & 0.0 &   5.5 & 2.41 & 2.42\\
        revDSD-PBEP86-D4 && 0.4224	& 0.5935 & 0.0566 & 0.5917 & 0.371 & 4.201 & 2.25 & 2.26\\
        revDOD-PBEP86-D4 && 0.4301 & 0.6131 & 0.0 & 0.6158  & 0.344 & 4.243 & 2.27 & 2.28\\
        \hline
        revDSD-PBEP86-D3BJ & VWN5 & 0.4310 & 0.5899 & 0.0727 & 0.4308  & 0.0 & 5.5 & 2.41 & 2.42\\
        revDOD-PBEP86-D3BJ && 0.4427 & 0.6155 & 0.0 & 0.4708  & 0.0 & 5.5 & 2.44 & 2.45\\
        revDSD-PBEP86-D4 && 0.4200 & 0.6090 & 0.0476 & 0.5850  & 0.370 & 4.183 & 2.29 & 2.30\\
        revDOD-PBEP86-D4 && 0.4278 & 0.6224	& 0.0 & 0.6134  & 0.348 & 4.242 & 2.31 & 2.31 \\
        \hline\hline
    \end{tabular}
    
    HF exchange $c_{\rm X,HF}$=0.69 throughout; $s_8$=0 throughout; $c_{\rm ATM}$=1.0 for D4 and not applicable for D3BJ.
    \caption{Optimized parameters and total WTMAD2(kcal/mol) of revDSD-PBEP86 functionals with PZ81 and VWN5 LDA correlations, respectively.}
    \label{tab:wtmad2}
    }
\end{table}

    
    The optimized parameter sets of revDSD-PBEP86 and revDOD-PBEP86 functionals with PZ81-P86c and VWN5-P86c are quite close to each other but not equivalent. The \markNM{difference between the WTMAD2 values} is within the uncertainty of the reference data of \markNM{the} GMTKN55 test suite (Table \ref{tab:wtmad2}). Moreover, swapping parameter sets between the two has an effect of less than 0.01 kcal/mol on WTMAD2.
    
    For the haVTZ+d and haVQZ+d basis sets, we evaluated analytical frequencies in three ways: 
    (a) with the original parameters of revDSD-PBEP86-D4, where the P86c is based on PZ81 local correlation; (b) with the same parameters, but using VWN5-based P86c instead; (c) with the purpose-optimized parameters, where the P86c of revDSD-PBEP86-D4 is based on VWN5. 
    Statistics for these three variants are shown in Table \ref{tab:freqrms} for haVQZ+d only, but the trends are the same for haVTZ+d. There are small but noticeable differences between the first and second variants of revDSD-PBEP86-D4 if PZ81 parameters are used for both, but the difference mostly goes away if the purpose-optimized parameters for VWN5 from Table \ref{tab:wtmad2} are used.

%
%

\section{Results and discussion}


\begin{table}[ht]
    \centering
\scriptsize{
    \begin{tabular}{c|ccc|c}
       RMSD w.r.t.  &  Experiment & CCSD(T*)(F12*)/ & CBS limit &MSD\\
                  &&  cc-pVQZ-F12 & of same method & w.r.t.Expt.\\
     \hline
        CCSD(T*)(F12*)/cc-pVQZ-F12 & 4.9 & REF &  & +0.3\\
        CCSD(T)/haV\{Q,5\}Z+d$^a$ & 4.4 & 0.8 &  & -0.1\\
        CCSD(T$_s$)(F12*)/cc-pVTZ-F12 & 4.7 & 0.8 && +0.1\\
        CCSD(T$_s$)(F12*)/cc-pVDZ-F12 & 5.7 & 3.3 && +0.2\\   
        CCSD(T)/haVDZ+d & 34.9 & 35.6 & & -25.7\\
        CCSD(T)/haVTZ+d & 11.1 & 11.5 & & -8.5\\
        CCSD(T)/haVQZ+d & 4.9 & 3.9 & & -2.7\\
        CCSD(T)/haV5Z+d & 4.4 & 1.9 & & -1.1\\
        CCSD(T)/haV6Z+d$^b$ & 3.6 & 1.7 & & $\Delta=0.6$\\
        \hline
        CCSD(T)/VDZ-F12$^c$ & 20.7 & 21.5 &  & -12.5\\
        CCSD(T)/VTZ-F12$^c$ & 10.6 & 10.4 &  & -8.4\\
        CCSD(T)/VQZ-F12$^c$ & 6.0 & 4.9 &  & -3.8\\
        \hline
        CCSD(F12*)/cc-pVQZ-F12 & 32.3 & 30.9 & REF& +26.2\\
        CCSD(F12*)/cc-pVTZ-F12 & 31.9 & 30.4 & 0.7 & +26.0\\
        CCSD(F12*)/cc-pVDZ-F12 & 31.1 & 29.7 & 2.2& +25.8\\
        CCSD-F12b/cc-pVDZ-F12 & 28.4 & 26.9 & 4.6& +23.2\\
         CCSD-F12b/cc-pVTZ-F12 & 29.9 & 28.5 & 2.5& +24.2\\
            CCSD-F12b/cc-pVQZ-F12 & 32.0 & 30.4 & 1.0& +26.1\\
        CCSD/haVDZ+d & 18.6 & 18.4 & 37.0 & -1.8\\
        CCSD/haVTZ+d & 21.9 & 20.5 & 12.0 & +17.6\\
        CCSD/haVQZ+d & 28.4 & 27.0 & 4.1 & +23.7\\
        CCSD/haV5Z+d & 31.1 & 29.8 & 2.0 & +25.4\\
       \hline
        MP2-F12/cc-pVQZ-F12 & 34.3 & 35.5 & REF & +11.0\\
        MP2-F12/cc-pVTZ-F12 & 33.3 & 34.0 & 0.5 & +11.4\\
        MP2-F12/cc-pVDZ-F12 & 34.1 & 34.8 & 2.3 & +10.8\\
        MP2/haVDZ+d & 45.9 & 47.0 & 24.8 & -7.8\\
        MP2/haVTZ+d & 34.0 & 35.0 & 8.20 & +5.8\\
        MP2/haVQZ+d & 33.8 & 34.6 & 3.0 & +10.1    \\
        MP2/haV5Z+d & 34.2 & 35.4 & 1.7 & +10.5 \\
        \hline\hline
    \end{tabular}
    \caption{RMS deviation (\wavenum) for HFREQ25 frequencies for various WFT methods from experiment (compiled in Refs.\cite{jmlm258,jmlm260}), from the CBS limit of the method being considered, and from CCSD(T)/CBS; mean signed deviation (\wavenum) from experiment}
    \label{tab:freqrmswft}
    
    $^a$ pointwise basis set extrapolation of energies ~~~\\
    $^b$ diatomics only ~~~\\
    $^c$ control run to gauge importance of F12 terms in isolation
    }
\end{table}

\subsection{WFT methods}

Let us first revisit wavefunction methods for some perspective before we consider the DFT methods (Table~\ref{tab:freqrmswft}.)
%
In Ref.\cite{jmlm258}, the best available calculated frequencies were either at the CCSD(T)/haV\{Q,5\}Z+d level --- pointwise extrapolation of energies to the complete basis set limit (CBS) at every point ---  or at the CCSD(T*)(F12*)/cc-pVQZ-F12 level, where the (T*) indicates that  we apply pointwise Marchetti-Werner\markNM{\cite{Marchetti-Werner}} scaling to the triples. (The triples do not benefit from F12.) Now strictly speaking (T*) is not size-consistent, but scaling the triples with a constant factor as proposed in Ref.\cite{jmlm261}, i.e. (Ts), definitely is size-consistent. (See Table 3 in that reference for numerical values of the scale factors.)

We recalculated harmonic frequencies at the CCSD(Ts)(F12*)/cc-pVQZ-F12 level for a subset of HFREQ25, and found an RMSD discrepancy between the harmonic frequencies of just 0.3 \wavenum — hence the  CCSD(T*)(F12*)/cc-pVQZ-F12 reference values from Ref.\cite{jmlm258} can be safely used.

For what it is worth, when we recalculated the smaller basis set frequencies at the CCSD(Ts)(F12*)/cc-pVTZ-F12 level, we found a surprisingly small RMSD=0.8 \wavenum, which increased to just 3.3 \wavenum when the small cc-pVDZ-F12 basis set was substituted instead. These RMSDs are nearly identical to the corresponding RMSDs for CCSD(T*)(F12*)/cc-pVnZ-F12 (n=T,D) from Table 1 in Ref.\cite{jmlm258}, i.e., 0.7 and 2.7 \wavenum, respectively.

Either CCSD(T*)(F12*)/cc-pVQZ-F12,  or CCSD(T)/aug-cc-pV(\{Q,5\}+d)Z with pointwise extrapolation,  can safely be considered to be within 1 \wavenum of the CCSD(T) basis set limit. 
As CCSD(T)/aug-cc-pV(\{Q,5\}+d)Z failed for CH$_3$OH but the results were available at the CCSD(T*)(F12*)/cc-pVQZ-F12 level, we used the latter as our CCSD(T)/CBS calibration point. It differs by 4.5 \wavenum from the `exact' reference harmonic frequencies used in Refs.\cite{jmlm258,jmlm260}{}: the residual difference is due to a combination of post-CCSD(T) correlation contributions, core-valence correlation, and scalar relativistic effects.\cite{jmlm117,jmlm230}

At the CCSD(F12*)/cc-pVQZ-F12 level, we find RMSD=32.3 \wavenum from `exact', and 30.9 \wavenum from CCSD(T), illustrating the great importance of connected triple excitations for especially stretching frequencies. (This has been known since the early 1990s at the latest, see, e.g., Refs.\cite{Rendell1990,jmlm042,jmlm069,Pawlowski2003}.) In fact, this performance is not much better than MP2-F12/cc-pVTZ-F12, which reaches 34.0 and 33.3 \wavenum, respectively.

CCSD(Ts)(F12*)/VDZ-F12 is actually comparable, if not superior, to CCSD(T)/haVQZ.

Recently Kruse et al.\cite{Kruse2020} suggested that cc-pVnZ-F12 basis sets might be superior to their haVnZ+d counterparts even in non-F12 calculations. As can be seen in Table \ref{tab:freqrms}, the statement holds true for CCSD(T)/VDZ-F12 vs. CCSD(T)/haVDZ+d, but already for VTZ-F12 and VQZ-F12 there is little daylight between these basis sets and their respective haVTZ+d and haVQZ+d counterparts. At the risk of belaboring the self-evident, comparing the RMSDs of CCSD(T)/VDZ-F12 and of CCSD(F12*)(Ts)/VDZ-F12 this illustrates that the introduction of the F12 geminals, not the basis set change as such, is what causes the basis set convergence acceleration.

As the (T) does not benefit from F12, we can eliminate a confounding factor by looking at CCSD.
Here we see that CCSD(F12*)/VDZ-F12 is actually superior to CCSD(T)/haVQZ+d,
and arguably closer to CCSD(T)/haV5Z+d in quality. One can say that, relative to the CBS limit, (F12*) gains three zetas
in the basis set, although it should be kept in mind that VDZ-F12 actually is considerably
larger than haVDZ+d. The popular CCSD-F12b approximation, on the other hand, lags CCSD(F12*)
by about one zeta step: see Ref.\cite{martin_opus_285} for a detailed analysis in the context of thermochemistry.

For MP2-F12 vs. conventional MP2, we again see a gain of 2.5-3 zeta steps, which looks promising for double hybrids.

Note that better basis set convergence does not {\em necessarily} translate into a lower RMSD with experiment.
CCSD at the CBS limit overestimates the reference by 26 \wavenum on average, and basis set incompleteness leads to
underestimated frequencies, so CCSD with a small basis set deceptively has a smaller RMSD from experiment owing to
error compensation. In contrast: at the CCSD(T) level, which at the CBS limit has very small RMSD (5 \wavenum) and negligible bias w.r.t. experiment (0.3 \wavenum),
improving the basis set consistently reduces RMSD w.r.t. experiment. The MP2 level takes a middle position: its bias is farly small (11 \wavenum) but
the RMSD relatively large (34 \wavenum) at the CBS limit. Here we see that the smallest basis set, haVDZ+d, is clearly inadequate, but already for haVTZ+d, RMSD w.r.t experiment reaches saturation.





%

\subsection{B2GP-PLYP double hybrid}

\begin{table}[ht]
    \centering
\scriptsize{
    \begin{tabular}{c|ccc|c}
       RMSD w.r.t.  &  Experiment & CCSD(T*)(F12*)/ & CBS limit &MSD\\
                    &&  cc-pVQZ-F12 & of same method & w.r.t.Expt.\\
        \hline
         revDSD-PBEP86-F12-D4/VQZ-F12 & 9.0 & 9.1 & REF &+4.0\\      
         revDSD-PBEP86-F12-D4/VTZ-F12 & 9.0 & 9.1 & 0.5 & +3.8\\ 
         revDSD-PBEP86-F12-D4/VDZ-F12 & 9.4 & 9.8 & 2.0 & +3.5\\   
         revDSD-PBEP86-F12-D4/haVDZ-F12 & 9.3 & 9.8 & 2.6 & +3.0\\   
        revDSD-PBEP86-D4/haVDZ+d$^c$ & 28.6 & 29.3 & 29.7 & -12.8\\     
           revDSD-PBEP86-D4/haVTZ+d$^c$ & 9.4 & 10.3 & 5.5 & +1.2\\     
        revDSD-PBEP86-D4/haVQZ+d$^c$ & 9.2 & 9.7 & 1.4 & +3.8 \\
        revDSD-PBEP86-D4/haVQZ+d$^a$  & 8.7 & 9.1 & 3.3 & +1.8 \\
        revDSD-PBEP86-D4/haVQZ+d$^d$ & 9.0 & 9.7 & 3.2 & +2.1 \\
         revDSD-PBEP86-D4/haVQZ+d$^e$ & 8.9 & 9.6 & 5.2 & +0.2 \\
        revDSD-PBEP86-D4/haV5Z+d$^b$ & 9.2 & 9.5 & 2.1 & +4.6\\
        ditto PZ81 instead of VWN5 & 8.5 & 8.8 & 2.4 & +2.7\\
        revDSD-PBEP86-D4/def2-TZVPP$^b$ & 10.8 & 11.1 & 4.4 & +6.2\\
             revDSD-PBEP86-D4/def2-QZVPP$^b$ & 8.9 & 9.8 & 2.2 & +5.4\\
                  revDSD-PBEP86-D4/def2-TZVPPD$^b$ & 9.0 & 10.5 & 3.5 & +4.0  \\
                       revDSD-PBEP86-D4/def2-QZVPPD$^b$ & 8.6 & 9.6 & 1.3 & +4.8\\
                                \hline
           revDSD-PBEP86-D4/VDZ-F12$^f$ & 11.8 & 14.8 & 11.5 & -0.3\\   
              revDSD-PBEP86-D4/VTZ-F12$^f$ & 7.9 & 9.3 & 4.9 & +1.0\\  
              revDSD-PBEP86-D4/VQZ-F12$^f$ & 8.1 & 9.0 & 2.3 & +3.0\\  
         \hline
        B2GP-PLYP-F12/VQZ-F12 & 23.0 & 23.1 & REF & +19.4\\      
        B2GP-PLYP-F12/AVTZ-F12 & 22.4 & 22.5 & 0.8 & +18.9\\    
        B2GP-PLYP-F12/VTZ-F12 & 22.6 & 22.7 & 0.5 & +19.1\\    
         B2GP-PLYP-F12/AVDZ-F12 & 22.7 & 23.0 & 2.2 & +18.6\\    
        B2GP-PLYP-F12/VDZ-F12 & 22.9 & 23.2 & 1.9 & +19.0\\    
        B2GP-PLYP-F12/haVDZ-F12 & 22.6 & 22.9 & 2.2 & +18.7\\  
        B2GP-PLYP-F12/haVTZ-F12 & 22.5 & 22.6 & 0.7 & +18.9\\  
        B2GP-PLYP/haV5Z+d & 22.6 & 22.6 & 0.9 & +19.3\\
        B2GP-PLYP/haVQZ+d & 22.4 & 22.7 & 1.6 & +18.7\\
        B2GP-PLYP/haVTZ+d & 20.7 & 21.1 & 4.5 & +16.6\\
        B2GP-PLYP/haVDZ+d & 26.7 & 27.4 & 27.4 & +3.4\\
        B2GP-PLYP/def2-QZVPP & 23.6 & 23.7 & 2.5 & +20.1\\
        B2GP-PLYP/def2-TZVPP & 24.6 & 24.8 & 4.9 & +21.3\\
        B2GP-PLYP/def2-QZVPPD & 23.3 & 23.4 & 1.7 & +19.5\\
        B2GP-PLYP/def2-TZVPPD & 23.5 & 23.7 & 3.5 & +19.2\\
        \hline
        $\omega$B97X-V/haVQZ & 39.8 & 38.3 & & 25.9\\
        $\omega$B97M-V/haVQZ & 34.3 & 33.0 & & 16.8\\
        $\omega$B97M(2)/haVQZ & 27.3 & 27.4& & 21.7\\
        \hline\hline
    \end{tabular}
    \caption{RMS deviation (\wavenum) for HFREQ25 frequencies from experiment (compiled in Refs.\cite{jmlm258,jmlm260}), from the CBS limit of the method being considered, and from CCSD(T)/CBS; mean signed deviation (\wavenum) from experiment}
    \label{tab:freqrms}

    $^a$ Gaussian 16, analytically, grid=superfine, PZ81 LDA \\
    $^b$ using VWN5 for LDA correlation part (PZ81 not available in MOLPRO)\\
    $^c$ Gaussian 16, analytically, grid=superfine, VWN5 LDA with PZ81-optimized parameters from Ref.\cite{revDSDfuncMartin} ~~~  \\
    $^d$ Gaussian 16, analytically, grid=superfine, VWN5 LDA with reoptimized parameters from Table~\ref{tab:wtmad2} ~~~ \\
        $^e$ Gaussian 16, analytically, grid=superfine, PZ81 LDA with VWN5-optimized parameters from Ref.\cite{revDSDfuncMartin} ~~~  \\
    $^f$ control run without F12 geminals, to gauge importance of F12 terms in isolation from the use of cc-pV$n$Z-F12 basis sets.
    }
\end{table}

For comparison, the B2GP-PLYP double hybrid with the aug-cc-pV(5+d)Z basis set clocks in at 22.6 \wavenum relative to either reference. (Table \ref{tab:freqrms}.)

For a subset of molecules, we were able to perform B2GP-PLYP/aug-cc-pV(6+d)Z frequency calculations, which differed by 0.6 \wavenum RMS from the smaller aug-cc-pV(5+d)Z. This number is not {\em that} much larger than the RMSD of 0.25 \wavenum between double-numerical (MOLPRO) and analytical (Gaussian) B2GP-PLYP/aug-cc-pV(Q+d)Z frequencies. Hence we can safely consider B2GP-PLYP/aug-cc-pV(5+d)Z to be within RMSD$<$1 \wavenum of the basis set limit, and 23 \wavenum to be a realistic estimate of the intrinsic error of the B2GP-PLYP functional.

B2GP-PLYP-F12/cc-pV$n$Z-F12 differs by just 1.8 \wavenum RMS for n=D, 1.0 \wavenum for n=T, and 0.9 \wavenum for n=Q. 

Indeed, considering the RMS difference of just 0.5 \wavenum between B2GP-PLYP-F12/cc-pVTZ-F12 and B2GP-PLYP-F12/cc-pVQZ-F12, it is plausible to assume that the latter numbers may well be {\em closer} to the CBS limit than B2GP-PLYP/haV5Z+d. 

 RMSDs from the CBS limit of B2GP-PLYP-F12/cc-pV(n-2)Z-F12 and B2GP-PLYP/haVnZ+d seem to be comparable: hence, we see that F12 gains about two zetas on average. Indeed, B2GP-PLYP-F12/cc-pVDZ-F12 at RMSD=1.9 \wavenum is not that much further from the cc-pVQZ-F12 numbers than B2GP-PLYP/haVQZ+d at 1.6 \wavenum. The informal `F12 gains two zetas in the basis set' adage seems to hold here as well.

But the important thing to realize here is that even for the modest \markNM{haVTZ+d} basis set, the RMSD relative to `exact', 21 \wavenum is no worse than 23 \wavenum, and indeed (fortuitously) somewhat better. Hence, unless the object is to eliminate basis set incompleteness error for calibration purposes, there is no real gain to be gotten here from VnZ-F12 calculations, as getting closer to the CBS limit has no benefits from a strictly spectroscopy point of view.

\subsection{revDSD-PBEP86-D4 double hybrid}

The situation for revDSD-PBEP86-D4 is somewhat different. RMSD from `exact' is just 9.0 \wavenum (!) at the revDSD-PBEP86-F12/cc-pVQZ-F12 level. These latter numbers are approximated to RMS=2.0 and 0.5 \wavenum, respectively, by the cc-pVDZ-F12 and cc-pVTZ-F12 basis sets.

Once again, DH-F12 gains about two zetas over orbital-only DH for this functional. Yet even at the fairly modest revDSD-PBEP86/haVTZ+d level, which differs 5.0 \wavenum RMS from the putative CBS limit, RMS relative to the best results increases to only 9.4 \wavenum. So again, we see that the RMSD with respect to experiment is approaching saturation with an haVTZ+d basis set, and that pushing further is only beneficial in the context of a calibration study. 

The benefit from including the F12 geminal can be shown by comparing RMSD for the VnZ-F12 basis sets with and without inclusion of the geminal terms. For the VDZ-F12 basis set, RMSD with respect to the CBS limit drops from 11.5 to 2.0 \wavenum upon introducing the F12 geminal, and for VTZ-F12 from 4.9 to 0.5 \wavenum. RMSD with respect to experiment decreases from 11.8 to 9.4 \wavenum for VDZ-F12, but (owing to error `decompensation') actually goes up from 7.9 to 9.0 \wavenum for VTZ-F12.


\subsection{Some observations on other functionals}

The combinatorially optimized $\omega$B97M(2) functional of Mardirossian and Head-Gordon\cite{wB97M2} slightly outperforms revDSD-PBEP86-D4 for the GMTKN55 benchmark suite. The mind wonders if the performance of the two functionals for vibrational frequencies would likewise be comparable. While we were at it, we added statistics for the fourth-rung ancestors of $\omega$B97M(2), namely, $\omega$B97X-V\cite{wB97XV} and $\omega$B97M-V.\cite{wB97MV} We did not carry out a basis set convergence study for these functionals but considered only the haVQZ basis set, deeming it close enough to the basis set limit for our purposes. 

First of all, for the three range-separated hybrids (RSH), dispersion-corrected `combinatorial Berkeley functionals', there is a clear reduction of RMSD from the RSH GGA $\omega$B97X-V via the RSH mGGA $\omega$B97M-V to the RS-double hybrid $\omega$B97M(2): 40 to 34 to 27 \wavenum. That said, the latter number is three times the 9 \wavenum for revDSD-PBEP86-D4 --- despite the fact that the latter only involves six adjustable parameters rather than sixteen, and that frequencies were not at all part of the training set. This gives us some confidence that the form of revDSD-PBEP86-D4 does not merely represent a lucky accident but compactly describes some real physics.

\section{Conclusions}

It has been previously shown\cite{jmlm314,jmlm318} for energetic properties that the introduction of F12 corrections in double-hybrid density functional theory drastically accelerates (by 2--3 zetas) basis set convergence and indeed allows near-basis set limit results already with the modest cc-pVDZ-F12 basis set. 

For vibrational frequencies, F12 still accelerates basis set convergence by about 2 zetas. However, basis set dependence for vibrational frequencies, while still significant, is weaker than for energetics, and the residual basis set incompleteness errors are much smaller (almost an order of magnitude even for the VDZ-F12 basis set) than the intrinsic error in the functional.

DHDF-F12 with a modest F12-optimized basis set may still be an attractive option to inexpensively establish a functional's CBS limit for calibration purposes. However, until analytical MP2-F12 derivatives with non-HF reference orbitals become available, conventional approaches with analytical second derivatives will still be preferable (albeit with a much larger memory footprint) for medium-large molecules. For small-medium molecules, the disadvantage of having \markNM{to evaluate} a large number of single-point energies is compensated to some degree by the `embarrassingly parallel' nature of the process, while even the best-parallelized analytical gradient code would not achieve 100\% parallelization efficiency.

If existing analytical derivatives\cite{Hofener2010,Gyorffy2017} for MP2-F12 could be adapted for double hybrids, then DH-F12 technique might still prove attractive for frequencies as it effectively eliminates concerns about inadequate basis set convergence.

For the accurate revDSD-PBEP86-D4 functional, the role of differing local correlation terms (Perdew-Zunger vs. VWN5) in different electronic structure programs has been investigated: while optimal double hybrid parameters and performance statistics for energetics as well as frequencies differ slightly between the two implementations, these differences are insignificant for practical purposes.

Finally, the good performance of $\omega$B97M(2) for energetic properties does not carry over to vibrational frequencies, unlike for revDSD-PBEP86-D4.


\begin{acknowledgement}
Work on this paper was supported in part by the Israel Science Foundation (grant 1969/20), by the Minerva Foundation (grant 2020/05), and by a research grant from the Artificial Intelligence and Smart Materials Research Fund, in memory of Dr. Uriel Arnon, Israel. \markNM{Two co-authors would like to acknowledge the following fellowships from the Feinberg Graduate School: Golokesh Santra a doctoral fellowship, and Nisha Mehta a Sir Charles Clore Postdoctoral Fellowship as well as Dean of the Faculty and Weizmann Postdoctoral Excellence fellowships.}

\end{acknowledgement}

\subsection*{Competing interests}

The authors report no competing interests.

\subsection*{Data availability statement}

The computed harmonic frequencies are gathered in the Supporting Information to the paper. Further raw data are available on reasonable request from the authors.

\markNM{A preprint of this paper has been posted on arXiv.org.\cite{jmlm321}}

\begin{suppinfo}
Spreadsheet in Microsoft Excel format with all calculated harmonic frequencies at the various levels of theory.
\end{suppinfo}

\bibliography{DH-F12-freq}

\end{document}